\documentstyle[twoside,fleqn,espcrc2]{article}

\newcommand{\AmS}{{\protect\the\textfont2
  A\kern-.1667em\lower.5ex\hbox{M}\kern-.125emS}}

\newcommand{\be}{\begin{equation}}
\newcommand{\ee}{\end{equation}}

\title{\normalsize \hfill UWThPh-2000-45 \\
\hfill November 2000\\ [1 cm] \Large
Elastic $\nu e^-$ scattering of solar neutrinos with
electromagnetic moments}

\author{W. Grimus\thanks{grimus@doppler.thp.univie.ac.at} and
\underline{T. Schwetz}\thanks{schwetz@doppler.thp.univie.ac.at.
T.S. is supported by the Austrian Academy of Science.}
\address{Institute for Theoretical Physics, University of Vienna,\\
       Boltzmanngasse 5, A-1090 Vienna, Austria}}

\begin{document}

\begin{abstract}
We consider the azimuthal asymmetry of the recoil electrons in
elastic $\nu e^-$ scattering of solar neutrinos, which can arise
if neutrinos have electromagnetic moments and there is a large solar magnetic
field. We show that using this effect it is
not possible to distinguish between magnetic and electric dipole moments
in the 1-Dirac and 2-Majorana neutrino cases and that averaging over
neutrino energy is important and suppresses the azimuthal asymmetry in the
2-Majorana neutrino case.
\end{abstract}

\maketitle

\footnote{Talk given by T.S. at EuroConference
on Frontiers in Particle Astrophysics and Cosmology, San Feliu de
Guixols, Spain, 30 Sept. - 5 Oct. 2000}
Neutrinos produced in the sun provide an interesting possibility
to investigate neutrino properties. Beside the well-established
neutrino oscillation search using solar neutrinos it is also well-known
that they can be used to look for a magnetic moment (MM)
or an electric dipole moment (EDM) of the neutrinos \cite{EMDM}.
If there exists a large magnetic field inside the sun
the so-called Resonant Spin-Flavour Precession scenario provides an
appealing sollution to the solar neutrino problem \cite{RSFP}.
Here we consider
elastic scattering of solar neutrinos with MMs and EDMs off electrons.
If the neutrinos acquire a transverse polarization because of 
the solar magnetic
field there can be an azimuthal asymmetry in the recoil electron
momentum \cite{asym}. Such an effect could be observed in a
$\nu e^-$ scattering experiment sensitiv to
low energy solar neutrinos with good angular resolution like the
proposed experiment HELLAZ \cite{HELLAZ}. Details of our considerations and
further references can be found in Ref.\cite{GS}.

{\bf The electromagnetic Hamiltonian.}
To describe the interaction of Dirac neutrinos with a MM and an EDM
with the electromagnetic field we use the Hamiltonian
\be
{\mathcal{H}}_{\mathrm{em}}^D =
\frac{1}{2} \bar{\nu}_R \lambda
\sigma^{\alpha \beta} \nu_L F_{\alpha \beta} + \mathrm{h.c.} 
\label{HD}
\ee 
Here
$\nu^T_{L(R)} = (\nu_e , \nu_\mu , \nu_\tau , \nu_s , \ldots )_{L(R)}$ 
is the vector of the left-handed (right-handed)
flavour eigenfields including an arbitrary number of sterile neutrinos.
The hermitian matrices $\mu$ of MMs and $d$ of EDMs
are condensed in the non-hermitian matrix
\be\label{lambda}
\lambda = \mu - i d \:\:
\mbox{with}\:\: \mu = \frac{ \lambda + \lambda^\dagger}{2} \,, \;
d = \frac{i( \lambda - \lambda^\dagger )}{2}\,.
\ee
If the basis of the neutrino fields is changed by unitary rotations
$\nu_{L} = S_L \nu'_L$ and $\nu_{R} = S_R \nu'_R$
the matrix (\ref{lambda}) in the new basis is obtained by
the simple relation
$\lambda' = S_R^\dagger \lambda S_L$,
whereas $\mu$ and $d$ obey rather complicated
transformation laws \cite{GS}.

Similarly, for Majorana neutrinos we have the Hamiltonian \cite{SV}
\be
{\mathcal{H}}_{\mathrm{em}}^M
= - \frac{1}{4} \nu_L^T C^{-1} \lambda
\sigma^{\alpha \beta} \nu_L F_{\alpha \beta} + \mathrm{h.c.} \,,
\label{HM}
\ee
where $C$ is the charge conjugation matrix.
Now the matrix $\lambda$, defined as in Eq.(\ref{lambda}),
is antisymmetric and the MM and EDM matrices are
antisymmetric and hermitian.

{\bf The cross section.}
In addition to the electromagnetic interaction there is also the
weak interaction of neutrinos with electrons, which
is described by the Hamiltonian
\be\label{HW}
{\mathcal{H}}_{\mathrm{w}} = \frac{G_F}{\sqrt{2}} \sum_\alpha
\bar{\nu}_\alpha \gamma_\lambda (1 - \gamma_5)\nu_\alpha \:
\bar{e} \gamma^\lambda (g_V^\alpha - g_A^\alpha \gamma_5) e \,,
\ee
where $G_F$ is the Fermi constant and
$g_V^\alpha =  2 \sin^2\Theta_W + d^\alpha,\,
g_A^\alpha = d^\alpha,\,
g_{V,A}^s =  0$,
with the weak mixing angle $\Theta_W$ and
$d^e = 1/2,\, d^{\mu,\tau} = -1/2$.

In the general case of an arbitrary neutrino polarization
the cross section for elastic $\nu e^-$ scattering consists
of three terms
\be\label{cross}
\frac{d^2 \sigma}{dT d \phi} = \frac{d^2 \sigma_{\mathrm{w}}}{dT d \phi} +
\frac{d^2 \sigma_{\mathrm{em}}}{dT d \phi} +
\frac{d^2 \sigma_{\mathrm{int}}}{dT d \phi} \,,
\ee
where $\phi$ is the azimuthal angle which is measured in the plane
orthogonal to the momentum of the initial neutrino and $T$ is the
recoil energy of the scattered electron.
The first and the second term are the pure weak and electromagnetic terms,
respectively, and the third term is the interference term between the weak
and the electromagnetic amplitude which
is proportional to the transverse neutrino polarization and
gives rise to the azimuthal asymmetry.

For the initial neutrino state in (\ref{cross})
we use an arbitrary superposition of
flavour and helicity states:
\be\label{instate}
|\nu\rangle_{\mathrm{in}} =
\sum_{\alpha=e,\mu,\tau,s,\ldots}
\left( a_-^\alpha |\nu^{(-)}_\alpha\rangle +
a_+^\alpha|\nu^{(+)}_\alpha\rangle \right)\,.
\ee
In the massless limit the negative helicity states are
left-handed neutrinos whereas
the positive helicity states are sterile right-handed neutrinos
in the Dirac case and right-handed antineutrinos
in the Majorana case.

The weak cross section for Majorana neutrinos is given by
\be\label{wcsm}
\frac{d^2 \sigma^M_{\mathrm{w}}}{dT d \phi} = \sum_\alpha \left(
|a_-^\alpha|^2 \frac{d^2 \sigma_{\nu_\alpha e}}{dT d \phi} +
|a_+^\alpha|^2
\frac{d^2 \sigma_{\bar{\nu}_\alpha e}}{dT d \phi} \right) \,,
\ee
where $\sigma_{\nu_\alpha e\,(\bar{\nu}_\alpha e)}$ is the
cross section for elastic scattering of neutrinos (antineutrinos)
of flavour $\alpha$ off electrons given e.g. in \cite{GS}. For
Dirac neutrinos the second term in Eq.(\ref{wcsm}) is absent.

The electromagnetic cross section has the same form for
Dirac and Majorana neutrinos:
\be\label{csEM}
\frac{d^2 \sigma_{\mathrm{em}}}{dT d\phi} =
c \left( \frac{1}{T} - \frac{1}{\omega} \right)
\left( a_-^\dagger \lambda^\dagger \lambda a_- +
a_+^\dagger \lambda \lambda^\dagger a_+ \right) \,,
\ee
where $c=\alpha^2/ 2m_e^2 \mu_B^2$,
$\omega$ denotes the neutrino energy and
$a_\mp^T = (a_\mp^e, a_\mp^\mu,\ldots)$.

The interference cross section is given by
\be
\frac{d^2 \sigma_{\mathrm{int}}}{dT d\phi} =
f\, \mbox{Re}\left[a_+^\dagger (\lambda g + \bar{g}\lambda)a_-
(p'_x - ip'_y)\right]\,. \label{csintM}
\ee
Here we have defined $f = G_F \alpha /2 \sqrt{2}\pi m_e T \mu_B$ and
$g = \mbox{diag}\left[ g_V^\alpha (2 - T/\omega) +
g_A^\alpha T/\omega \right]$. For Majorana neutrinos
$\bar{g} = \mbox{diag}\left[ g_V^\alpha (2 - T/\omega) -
g_A^\alpha T/\omega \right]$ whereas
$\bar{g}=0$ in the Dirac case.
For the direction of the initial neutrino we choose the $z$-axis and
the transversal components of the recoil electron momentum are related to
the azimuthal angle via $p'_x = p'_\perp \cos\phi,\,
p'_y = p'_\perp \sin\phi$ with ${p'_\perp}^2 = {p'_x}^2+{p'_y}^2$.
Therefore we find from Eq.(\ref{csintM})
\be\label{csint2}
\frac{d^2 \sigma_{\mathrm{int}}}{dT d\phi} \propto
\cos(\phi-\gamma)\,.
\ee
The measurement of the azimuthal asymmetry in an experiment would allow
to determine the angle $\gamma$ which is defined as
$\gamma \equiv \mbox{Arg}[a_+^\dagger (\lambda g + \bar{g}\lambda)a_-]$.

In the following we will consider the question:
{\it Is it possible to
obtain information on complex phases in the electromagnetic moment matrix
$\lambda$ or in the neutrino mixing matrix via a measurement of $\gamma$?}

{\bf Neutrino evolution in the sun.}
The evolution of the neutrino
state produced in the core of the sun
under the influence of the solar magnetic field and matter effects
is governed by the Schr\"odinger-like equation \cite{evol}
\begin{eqnarray}
 i\frac{d}{dz} \left(\begin{array}{c} \varphi_- \\ \varphi_+
\end{array}\right) = H_{\mathrm{eff}}
\left(\begin{array}{c} \varphi_- \\ \varphi_+  \end{array}\right)
\quad\mbox{with}\quad
\label{evol}\\
H_{\mathrm{eff}} \equiv
\left(\begin{array}{cc}
V_L + U_L\frac{\hat{m}^2}{2\omega}U_L^\dagger & -B_+ \lambda^\dagger \\
-B_-\lambda & V_R +  U_R\frac{\hat{m}^2}{2\omega}U_R^\dagger
\end{array}\right)\,. \nonumber
\end{eqnarray}
In this equation, $\varphi_-$  and $\varphi_+$
denote the vectors of neutrino flavour
wave functions corresponding to negative and positive helicity,
respectively.
$V_L = \sqrt{2} G_F \mbox{diag}(n_e - n_n/2, -n_n/2, -n_n/2, 0, \ldots)$
is the matter potential
where $n_e\:(n_n)$ is the electron (neutron) density in the
sun and $V_R=0\,(-V_L)$ for Dirac (Majorana) neutrinos.
The diagonal matrix of neutrino masses is denoted by $\hat{m}$ 
and $U_L$ is the unitary mixing matrix connecting left-handed 
flavour and mass eigenfields. $U_R$ is an arbitrary unitary matrix
for Dirac neutrinos and $U_R=U_L^*$ in the Majorana case.
Finally, $B_\pm = B_x \pm iB_y$ where $B_x$ and $B_y$ are the components
of the solar magnetic field orthogonal to the neutrino momentum.

Neutrinos are produced as electron neutrinos in the sun at the
coordinate $z_0$ and are detected on earth at $z_1$. Hence we express
the initial condition as
$\varphi_-^T(z_0) = (1,0,\ldots),\:
\varphi_+^T(z_0) = (0,\ldots)$
and for the neutrino state at the detector, Eq.(\ref{instate}),
we have $a_\mp \equiv \varphi_\mp(z_1)$.
For a given magnetic field along the neutrino path in the sun,
the neutrino state described by the vectors $a_\mp$ can in principle
be obtained by solving Eq.(\ref{evol}), as a function of neutrino MMs,
EDMs, masses and mixing parameters. 
These flavour vectors $a_\mp$ have to be used in the cross section
for elastic $\nu e^-$ scattering of solar neutrinos.

{\bf Phase counting.} Let us first consider a single Dirac neutrino.
In this simplest case
there is only one complex phase in the problem,
which is the phase $\delta$ defined through
$\mu + i d = \sqrt{\mu^2 + d^2}\, e^{i\delta}$ where $\mu$ and $d$ are
real numbers in this case. 
Now, the evolution Eq.(\ref{evol}) leads to
$a'_+ = e^{i\delta} a_+$ such that
$a_-$ and $a'_+$ as well as the angle $\gamma$ in the interference cross
section (\ref{csint2}) depend only on $\sqrt{\mu^2 + d^2}$.
Hence, the phase $\delta$ disappears from the problem.
This means that it is not possible to distinguish between a Dirac MM and
EDM via a measurement of $\gamma$. A nonzero $d$ leads
to CP violation at the level of the Lagrangian. However, the above
consideration implies that no CP violating effects can be observed in
elastic $\nu e^-$ scattering of solar neutrinos.

In the second simple case of two Majorana neutrinos there are two phases
in the problem: one in the transition moment $\mu - id$ in the matrix
$\lambda$ ($\mu$ and $d$ imaginary) and one phase in the
mixing matrix (Majorana phase). One can show with arguments similar to the
one Dirac case that both phases can be absorbed through redefinitions.
Again it is not possible to distinguish between a MM and an EDM and no
CP violating effects show up.

\begin{table}
\caption{Number of complex phases for $n$ flavours:}
\label{tab}
\begin{tabular}{l|cc}
\hline
& Dirac & Majorana \\
\hline
mix. matrix & $(n-1)(n-2)/2$ & $n(n-1)/2$ \\
$\lambda$ & $n^2$ & $n(n-1)/2$ \\
\hline
azim. asym. & $(3n-2)(n-1)/2$ & $n(n-2)$ \\
\hline
\end{tabular}
\end{table}

Now we come to the general case of $n$ neutrino flavours. In the first
two lines of Table \ref{tab} we give the numbers of complex phases
in the mixing matrix and in $\lambda$ at the level of the Lagrangian,
in a phase convention where as much phases as possible are removed
from the mixing matrix.
However, not all of these phases show up in elastic $\nu e^-$ scattering
of solar neutrinos because of the relevant physical approximations:
neutrino masses enter in the evolution equation (\ref{evol})
only via the terms
$ U_L \hat{m}^2 U_L^\dagger,\,  U_R \hat{m}^2 U_R^\dagger$ and
are neglected in the cross section. Therefore, in the physical situation
under consideration there is more phase freedom which can be used to
reduce the number of phases in the problem. Moreover, complex phases
can be shifted from the mixing matrix to $\lambda$ and vice versa \cite{GS}.
In the last line of Table \ref{tab} we show the total number of physical
phases relevant for the azimuthal asymmetry in elastic $\nu e^-$ scattering
of solar neutrinos.

{\bf Decoherence effects.}
Here we consider the effect of neutrino oscillations and
averaging over the neutrino energy
on the solar neutrino state arriving at the earth.

The neutrino state undergoes vacuum oscillations between
the sun and the earth. Therefore, denoting the values of 
$\varphi_\mp$ (\ref{evol}) at the edge of the sun by
$b_\mp$, we have
$a_- = U_L \exp \left( -i \hat{m}^2 L/2\omega \right) U_L^\dagger\, b_-$
and
$a_+ = U_R \exp \left( -i \hat{m}^2 L/2\omega \right) U_R^\dagger\, b_+$.
Here $L\approx 1.5 \times 10^{11}$ m is the distance between the sun and the
earth. Now the crucial point is that, according to the quadratic appearance
of $a_\mp$ in the cross section, the phase factors
$e^{\pm i\varphi_{jk}}$ are important with
$\varphi_{jk} = 2\pi L / \ell_{jk}$
where 
$\ell_{jk} = 4\pi \omega / \Delta m^2_{jk}$ is an oscillation length
with $\Delta m^2_{jk} = m^2_j - m^2_k > 0$.
The phases vary with energy as
\be
\delta \varphi_{jk} = \frac{\Delta m^2_{jk} L}{2\omega}
\frac{\delta\omega}{\omega}  =
2\pi \frac{L}{\ell_{jk}} \frac{\delta\omega}{\omega} \,.
\ee
Hence, integration over energy intervals such
$\delta\omega \gg  \omega \,\ell_{jk} / L$ $\forall j,k$
leads to an averaging of the oscillations, which can formally be
expressed as
$\left\langle e^{\pm i\varphi_{jk}} \right\rangle = \delta_{jk}$,
where $\delta_{jk}$ is the Kronecker delta.
Numerically, we have
\be\label{cohnum}
\frac{\ell_{jk}}{L}
\approx 1.7 \times 10^{-11} \: \frac{\omega (\mbox{MeV})}
{\Delta m^2_{jk} (\mbox{eV}^2)}\,.
\ee
If we consider, for example,
$\Delta m^2 \sim 10^{-8}$ eV$^2$ allowed by the RSFP scenario
\cite{RSFP} and
$\omega \approx 0.27$ MeV,
the average energy of the
$pp$ neutrinos, we find $\ell /L \sim 5 \times 10^{-4}$, where
$\ell$ is the oscillation length corresponding to $\Delta m^2$. Therefore,
to avoid the averaging
one would have to measure the neutrino energy with an accuracy better than
$\delta \omega /\omega \sim 10^{-4}$, which seems rather impossible.
The energy averaging of the vacuum
oscillations is equivalent to consider the neutrino
state arriving at the earth as an incoherent mixture of mass eigenstates.

The expressions for the general energy averaged cross sections are given
in Ref.\cite{GS}. Here we want to discuss the effect of decoherence for
the 2-Majorana neutrino case. For total incoherence the
interference cross section is proportional to
\be\label{csint2maj}
\left\langle \frac{d^2 \sigma^M_{\mathrm{int}}}{dT d\phi} \right\rangle
\propto f \,  \frac{T}{\omega}\, \sin 2\theta \cos(\phi-\gamma)\,
\ee
where $f$ is given after Eq.(\ref{csintM}).
Again all complex phases disappear,
especially $\gamma$ does not depend on the phase of $\mu - id$.
There are two important implications of relation (\ref{csint2maj}):
({\it i}) The dependence on the electron recoil energy of this expression
is very different from the corresponding term
in the case of full coherence and the Dirac terms with and without coherence,
because the recoil energy $T$ drops out of the product $f T$. (This is
also true for an arbitrary number of Majorana neutrinos.)
({\it ii}) Expression (\ref{csint2maj}) is proportional to the mixing angle
$\sin 2\theta$. Large values for $\sin 2\theta$ are disfavoured in the RSFP
scenario \cite{RSFP} and by the
non-observation of electron antineutrinos in Super-Kamiokande 
\cite{SKanu} and  hence the asymmetry is suppressed.

These arguments suggest that a significant asymmetry measured
in an experiment is unlikely to result from a 2-Majorana neutrino
scenario, except for very small mass-squared differences
($\Delta m^2 < 10^{-11}$ eV$^2$). Of course, it could
result from Dirac diagonal moments. In this case
the states of negative and positive helicity belong to the same
mass eigenvalue and no averaging due to oscillations is possible.

{\bf Conclusions.}
We have considered the possibility to investigate neutrino properties
using the azimuthal asymmetry in elastic $\nu e^-$ scattering of solar
neutrinos. We have shown that it is not possible to distinguish MMs and
EDMs for 1-Dirac and 2-Majorana neutrinos and no CP violation will show
up in these cases. For $n$ neutrino flavours there is a phase freedom
because of the physically motivated approximations. This allows
to eliminate complex phases in the mixing matrix and the MM/EDM matrix.
Furthermore we have shown that energy averaging is important and leads to
a suppression of the azimuthal asymmetry in the 2-Majorana case.

\end{document}